\documentclass[12pt,reqno,a4paper]{amsart}

\usepackage{lipsum,subeqnarray}
\usepackage{amsfonts}
\usepackage{graphicx}
\usepackage{amsmath}
\usepackage{mathtools}
\usepackage{amssymb}
\usepackage{enumerate}
\usepackage{slashed}
\usepackage{graphicx}
\usepackage{newlfont}
\usepackage{amsrefs}
\usepackage{comment}
\usepackage[normalem]{ulem}
\usepackage{mathtools}
\usepackage{accents}
\usepackage{leftidx}
\usepackage{bbm}
\usepackage{amsthm}

\numberwithin{equation}{section}

\theoremstyle{plain}
\newtheorem{theorem}{Theorem}[section]

\newtheorem{proposition}[theorem]{Proposition}
\newtheorem*{theorem*}{Conjecture}

\usepackage{xcolor}

\theoremstyle{definition}

\theoremstyle{remark}

\usepackage{stackengine}
\usepackage{subeqnarray} 

\stackMath
\newcommand\tenq[2][1]{%
 \def\useanchorwidth{T}%
  \ifnum#1>1%
    \stackunder[0pt]{\tenq[\numexpr#1-1\relax]{#2}}{\scriptscriptstyle\sim}%
  \else%
    \stackunder[1pt]{#2}{\scriptscriptstyle\sim}%
  \fi%
}

\newcommand{\overbar}[1]{\mkern 1.6mu\overline{\mkern-1.6mu#1\mkern-1.6mu}\mkern 1.6mu}

\begin{document}
\title[Nonexistence of Solutions to the GJE/Zero Divergence System]{Nonexistence of Solutions to the Coupled Generalized Jang Equation/Zero Divergence System}
 
\author{Jaroslaw S. Jaracz}

\begin{abstract}
In \cite{BrayKhuriPDE}, Bray and Khuri proposed coupling the generalized Jang equation to several different auxiliary equations. The solutions to these coupled systems would then imply the Penrose inequality. One of these involves coupling the generalized Jang equation to $\overline{div}(\phi q)=0$, as this would guarantee the non-negativity of the scalar curvature in the Jang surface. This coupled system of equations has not received much attention, and we investigate it's solvability. We prove that there exists a spherically symmetric initial data set for the Einstein equations for which there do not exist smooth radial solutions to the system having the appropriate asymptotics for application to the Penrose inequality.    
\end{abstract}
\maketitle

\section{Introduction and Statement of Results}\label{SEC:Intro}

\subsection{The Penrose Conjecture}

The Penrose inequality has been one of the most famous open conjectures in mathematical general relativity. Conjectured by Roger Penrose in the 1970's using a heuristic argument based on the established view point of gravitational collapse \cite{Penrose}, it relates the total mass $m$ of a spacetime to the surface area $A$ of a black hole in the spacetime via the inequality
\begin{equation}
    m \geq \sqrt{\frac{A}{16 \pi }}.
\end{equation}

A special case, known as the Riemannian Penrose inequality, was proven in the late 1990's for an asymptotically flat initial data set $(M,g)$ by Huisken and Illmanen using a weak version of the inverse mean curvature flow (IMCF) \cite{HuiskenIlmanen}, and independently by Hubert Bray using a conformal flow of metrics \cite{Bray}. In these cases, the black hole is represented by a minimal surface and the initial data set must have non-negative scalar curvature $R\geq 0$ (or to satisfy some assumptions which imply this condition). The $m$ in these cases is given by the ADM energy. For the definitions of the ADM energy and ADM mass, see \eqref{ADMEnergy} and \eqref{ADMMass}.

 The Penrose inequality for a general asymptotically flat initial data set $(M, g, k)$ where $k$ is the extrinsic curvature remains an open problem. It had previously been proven in the case of spherical symmetry where $m$ is given by the ADM energy, assuming the so-called dominant energy condition and where the black hole is mathematically represented by an outermost future or past apparent horizon \cite{Hayward}, with a different proof further establishing rigidity for the inequality given in \cite{BrayKhuri} using a Jang equation approach. The inequality has been extended to include charge and angular momentum and proven under certain conditions, but all of these require some kind of non-negativity for the scalar curvature. See for example \cites{KhuriDisconzi, KhuriWeinsteinYamada1, KhuriSokolowskyWeinstein, JaraczCylindricalPenrose }.

A popular mathematically precise formulation of the Penrose conjecture is the following:
\begin{theorem*}[Penrose Inequality] 
Let $(M, g, k)$ be an asymptotically flat initial data set satisfying appropriate fall-off conditions and the dominant energy condition with boundary $\partial M$ consisting of an outermost apparent horizon, with possibly multiple components. Let $N$ be any component of the boundary and let $A=A_{min}(N)$ denote the outermost minimal area enclosure of the component $N$. Then
\begin{equation}
    E_{ADM}\geq \sqrt{\frac{A}{16\pi}}
\end{equation}
where $E_{ADM}$ is the ADM energy. 
\end{theorem*}

Here apparent horizons play the roles of black holes. There is also a formulation where the ADM energy is replaced by the ADM mass. We mention that under weaker energy conditions the Penrose inequality does not hold. One of the key ingredients for the Penrose inequality is the black hole are law, which holds under the weaker assumption of the weak energy condition. However, in this setting there exists a counter-example satisfying all the assumptions of the above conjecture, with the dominant energy condtion replaced by the weak energy condition, see \cite{JaraczWEC}.

\subsection{The Generalized Jang Equation} 

Here and for the remainder of the paper, given some arbitrary function $f$, we write
\begin{equation*}
    \partial_{x^i}f = \partial_i f = f_i
\end{equation*}
for the partial derivatives. Also, if $f=f(r)$ is a function of a single variable, we write $f' = f_r $ when convenient. 

The Jang equation was proposed by Jang in \cite{Jang} and was successfully used by Schoen and Yau in \cite{SchoenYau2} to prove the positive energy theorem assuming the dominant energy condition. The idea of the proof was to solve the Jang equation which would give a hypersurface $\Sigma$ in the product manifold $(M\times \mathbb{R}, g+ dt^2)$ as the graph of a certain function $f$ and, after a further conformal deformation, this surface would have positive scalar curvature and the same ADM energy as the original data set. One could then apply the positive energy theorem which had been proven in the case of positive scalar curvature \cite{SchoenYau}.  

One might hope that having established the Riemannian Penrose inequality the same approach could be used. However, it was pointed out that the original Jang equation was not suited for this, for a variety of reasons, but nevertheless it remained a tantalizing idea. Hence, in \cite{BrayKhuriPDE}, Bray and Khuri proposed a modification of the Jang equation, presented below. The original Jang equation is just the special case of $\phi=1$.

Given an initial data set $(M, g, k)$, one looks for the hypersurface $\Sigma$, referred to as the Jang surface, given by the graph $t=f(x)$ inside the warped product space $(M\times \mathbb{R}, g+\phi^2 dt^2)$. One then looks for a surface satisfying the equation, called the generalized Jang equation
\begin{equation} 
    H_{\Sigma}-Tr_{\Sigma} K =0
\end{equation}
where $H_{\Sigma}$ is the mean curvature of $\Sigma$ and $Tr_{\Sigma} K$ denotes the trace of $K$ over $\Sigma$. Here $K$ is a nontrivial extension of the initial data. Letting $\partial_{x^i}=\partial_i$ for $1\leq i \leq 3$ and $\partial_{x^4}=\partial_t$, the extension is given by
\begin{align} \begin{split}
    K(\partial_i, \partial_{j})= K(\partial_{j}, \partial_{i})=k(\partial_{j}, \partial_{i}) \quad \text{for} \quad 1\leq i, j \leq 3 \\ 
     K(\partial_{i}, \partial_t)= K(\partial_{t}, \partial_{i})=0 \quad \text{for} \quad 1\leq i \leq 3 \\
     K(\partial_t, \partial_t)=\frac{\phi^2 g(\nabla f, \nabla \phi)}{\sqrt{1+\phi^2 |\nabla f|^2}} \end{split}
\end{align}
where $x^i, i=1, 2, 3$ are local coordinates on $M$. In local coordinates, the generalized Jang equation takes the form
\begin{equation} \label{GJE}
    \left( g^{ij} - \frac{\phi^2 f^i f^j}{1+\phi^2 |\nabla f |^2}  \right)\left( \frac{\phi \nabla_{ij} f +\phi_i f_j + \phi_j f_i}{\sqrt{1+\phi^2 |\nabla f|^2}} - k_{ij}
    \right)=0
\end{equation}
where $f^{j}=g^{ij} f_i$.

The tangent vectors to $\Sigma$ are given by 
\begin{equation*}
    X_i=\partial_i + f_i \partial_t
\end{equation*}
and hence the induced metric on $\Sigma$ is given by
\begin{equation}\label{Metricf}
    \bar{g}=g+\phi^2 df^2.
\end{equation}
The inverse metric is given by
\begin{equation} \label{InverseMetric}
    \bar{g}^{ij}=g^{ij} - \frac{\phi^2 f^i f^j}{1+\phi^2 |\nabla f |^2}.
\end{equation}

The unit normal to $\Sigma$ is given by
\begin{equation*}
    N=\frac{\nabla f - \phi^{-2}\partial_t}{\sqrt{\phi^{-2}+|\nabla f|_g^2}}
\end{equation*}
where $\nabla$ denotes the covariant derivative with respect to the $g$ metric.

We denote by $A$ the second fundamental form of $\Sigma$ in $M\times \mathbb{R}$ and by $\overbar{div}$ the divergence operator with respect to $\bar{g}$. We also define the $1$-forms $q$ and $w$ by 
\begin{equation}
    w_i = \frac{\phi f_i}{\sqrt{1+\phi^2 |\nabla f|^2}}, \quad 
    q_i=\frac{\phi f^j}{\sqrt{1+\phi^2 |\nabla f|^2}} \left( A_{ij}-(K\vert_{\Sigma})_{ij} \right),
\end{equation}
where $K\vert_\Sigma$ is the restriction of $K$ to $\Sigma$. One then has the following key formula for the scalar curvature of $(\Sigma, \bar{g})$:
\begin{equation} \label{ScalarCurvatureJang}
    \overline{R}=2\left( \mu - J(w)  \right) + \vert A-K\vert_{\Sigma} |^2 + 2|q|^2-2\phi^{-1} \overline{div}(\phi q).
\end{equation}
This is known as the generalized Schoen-Yau identity, see \cites{BrayKhuriPDE, BrayKhuri}. In the case where $\overbar{R}\geq 0$, the Riemannian Penrose inequaltity can be applied to the Jang surface.

\subsection{The coupled system and the main theorem}

Assuming the dominant energy condition, all the terms in \eqref{ScalarCurvatureJang} are non-negative, except possibly the last one. This then naturally leads to the coupled system of equations
\begin{align} \label{CoupledSystem}
    \begin{split}
         \left( g^{ij} - \frac{\phi^2 f^i f^j}{1+\phi^2 |\nabla f |^2}  \right)\left( \frac{\phi \nabla_{ij} f +\phi_i f_j + \phi_j f_i}{\sqrt{1+\phi^2 |\nabla f|^2}} - k_{ij}
    \right)&=0 \\
    \overline{div}(\phi q)&=0.
    \end{split}
\end{align}
Sine $q$ depends on $f$ and $\phi$, this is a system of two equations in those two unknown functions. In addition, even though the second equation turns out to be third order in $f$, for a fixed smooth $f$ it can be viewed as a degenerate second order elliptic equation in $\phi$, which gives some hope that the system might be solvable. As discussed in \cite{HanKhuri}, for application to the Penrose inequality one needs $\phi>0$ outside of the boundary, with $\phi$ potentially vanishing at the boundary. However, we have the following theorem.

\begin{theorem}\label{MainTheorem}
There exists a smooth, spherically symmetric, assymptotically flat initial data set $(M, g, k)$ satisfying the dominant energy condition and with boundary consisting of a compact outermost apparent horizon for which \eqref{CoupledSystem} does not possess any smooth radial solutions with $\phi>0$ outside of the boundary, and with the appropriate asymptotics for application to the Penrose inequality. \end{theorem}

Now, this leaves open the highly unlikely possibility that there exist non-radial solutions to \eqref{CoupledSystem} for the spherically symmetric initial data set given in Theorem \ref{MainTheorem}. This is unlikely as symmetries make solving differential equations easier, and hence if \eqref{CoupledSystem} possessed solutions with the appropriate properties, at least some of them should be radial, which the above shows is impossible. Nevertheless, since the equations are non-linear proving this is non-trivial. Hence, we formulate this as a conjecture, which we expect to resolve in a future paper. 

\begin{theorem*}
    If a spherically symmetric initial data set $(M, g, k)$ possesses solutions to \eqref{CoupledSystem} with the appropriate asymptotics for application to the Penrose inequality, at least some of them must be radial. Hence the initial data set of Theorem \ref{MainTheorem} possesses no such solutions, and thus this approach cannot be used to prove the Penrose inequality.
\end{theorem*}

We also mention that \eqref{CoupledSystem} was recently investigated in the case where $\overline{div}(\phi q)$ was linearized. In this case, under some very restrictive assumptions on the initial data, solutions were shown to exist, though of course this doesn't prove the Penrose inequality in general \cite{Williams}. 

Our paper grew out of attempting to solve \eqref{CoupledSystem} in the general case. The aim of Theorem \ref{MainTheorem} and of the above conjecture is to settle whether or not this is a viable approach to the Penrose inequality, so that researchers do not waste precious resources on a hopeless approach.

\subsection{A closer look at the divergence term}

Since $q$ is a $1$-form, the divergence is interpreted in the usual way of first raising the index to obtain a vector and then taking the divergence. In abstract index notation we have
\begin{equation*}
 \overbar{div}(\phi q)=\overbar{\nabla}_a ( \phi \bar{g}^{ab}  q_b )
\end{equation*}
Recall, the formula for the divergence of a vector field with respect to the metric $g$ in local coordinates is given by
\begin{equation*}
    div_g(V) = \frac{1}{\sqrt{\vert {g} \vert}} \partial_k \left( \sqrt{ \vert {g} \vert } V^k \right) 
\end{equation*}
so we can rewrite 
\begin{align} \begin{split}
    \overbar{div}(\phi q)&= \frac{1}{\sqrt{\vert \bar{g} \vert}} \partial_k \left( \sqrt{ \vert \bar{g} \vert } \phi \bar{g}^{ki} q_i \right) \\ &= \frac{1}{\sqrt{\vert \bar{g} \vert}} \partial_k \left( \sqrt{ \vert \bar{g} \vert } \phi \bar{g}^{ki} \frac{\phi f^j}{\sqrt{1+\phi^2 |\nabla f|^2}} \left( A_{ij}-(K\vert_{\Sigma})_{ij} \right) \right). \end{split}
\end{align}
The components of $A$ are given by
\begin{equation*}
    A_{ij}=\left\langle \widetilde{\nabla}_{X_i} N, X_j \right\rangle_{\widetilde{g}}=\frac{\phi  \nabla_{ij} f + \phi_i f_j +\phi_j f_i + \phi^2 f^m \phi_m f_i f_j}{\sqrt{1+\phi^2 | \nabla f|_g^2}}
\end{equation*}
where $\nabla_{ij} f$ are the components of the second covariant derivative of $f$ with respect to the $g$ metric. On the other hand, we have
\begin{equation*}
    \left( \left.  K\right\vert_{\Sigma}\right)_{ij}=K(X_i, X_j)=k_{ij} + \frac{\phi^2  f^m \phi_m f_i f_j}{\sqrt{1+\phi^2 |\nabla f |_g^2}}
\end{equation*}
and so
\begin{equation*}
    A_{ij}-\left( \left.  K\right\vert_{\Sigma}\right)_{ij}=\left\langle \widetilde{\nabla}_{X_i} N, X_j \right\rangle_{\widetilde{g}}=\frac{\phi  \nabla_{ij} f + \phi_i f_j +\phi_j f_i}{\sqrt{1+\phi^2 | \nabla f|_g^2}} - k_{ij}
\end{equation*}
and therefore in local coordinates
\begin{align} 
     \overbar{div}(\phi q)=  \frac{1}{\sqrt{|\bar{g}|}}\partial_k \left( \frac{ \sqrt{|\bar{g}|} \phi^2 \bar{g}^{ki}f^j\left( \phi  \nabla_{ij} f + \phi_i f_j +\phi_j f_i \right)   }{1+\phi^2 |\nabla f |_g^2      }-\frac{ \sqrt{|\bar{g}|}\phi^2 \bar{g}^{ki} f^j k_{ij}  }{ \sqrt{1+\phi^2 |\nabla f|_g^2}  } \right).
     \end{align}

\section{Proof of Theorem \ref{MainTheorem}}

The initial data set we construct will be asymptotically flat and spherically symmetric. We begin by giving the appropriate definitions, then looking at the generalized Jang equation in spherical symmetry, coupling it to the zero divergence term, and analyzing the resulting ODE. We then show that the solutions to this ODE yield a contradiction.

\subsection{Asymptotic Flatness and the ADM Formalism} We consider an asymptotically flat initial data set $(M, g, k)$ where $M$ is a $3$-manifold, $g$ a Riemannian metric, and $k$ is a symmetric $2$-tensor, the extrinsic curvature. See \cites{ADM, Bartnik} for precise definitions. In each asymptotically flat end $g$ and $k$ satisfy certain fall-off conditions such as
\begin{align} \label{GeneralFallOffConditions} 
    | D^{\lambda} ( g_{ij} - \delta_{ij}) | \leq Cr^{-1-|\lambda|}, \quad |R|\leq Cr^{-3}, \quad |k|\leq Cr^{-2}, \quad |Tr_g k | \leq Cr^{-2} 
\end{align}
for $|\lambda|\leq 2$ which are standard. Here, $\delta$ is the Euclidean metric, $r=\sqrt{x^2+y^2+z^2}$ the standard Euclidean radius, $D^\lambda$ is a derivative operator with respect to the Euclidean coordinates, and $\lambda$ is a multi-index. We have 
\begin{equation*}
    |k|^2=k_{ij}k^{ij}, \quad Tr_g k = g^{ij}k_{ij}
\end{equation*}
as usual.  

For an asymptotically flat end, the ADM energy and ADM momentum are defined by 
\begin{align} \label{ADMEnergy} 
E_{ADM}=\lim_{r\rightarrow \infty}\frac{1}{16\pi} \sum_{i, j} \int_{S_r} (g_{ij,i}-g_{ii,j})\nu^j dS_r \\ \label{ADMMomentum}
P_i = \lim_{r\rightarrow \infty} \frac{1}{8\pi}  \sum_j \int_{S_r} \left(  k_{ji}\nu^j - (Tr_g k) \nu_i \right) dS_r
\end{align}
where $S_r$ are coordinate spheres of radius $r$ and $\nu^j$ is the outward unit normal \cite{ADM}. 
One then defines the ADM mass by
\begin{equation} \label{ADMMass}
    m_{ADM}=\sqrt{ E^2_{ADM} - |P|^2  }.
\end{equation}
It is well known \cites{Bartnik, ChruscielADM} that with our fall-off conditions these quantities do not depend on the choice of asymptotically flat coordinates.

\subsection{Energy Conditions and Constraint Equations}

For a full discussion of energy conditions in a $3+1$ spacetime $(\mathcal{M}, \mathfrak{g})$ we refer the reader to \cites{HawkingEllis, Wald}. The Einstein constraint equations for an initial data set $(M, g, k)$ are
\begin{align} \label{ConstraintEquations}
\begin{split}
16\pi\mu=R+(Tr_gk)^2-|k|^2\\
8\pi J_i=\nabla^j (k_{ij}-(Tr_g k)g_{ij})
\end{split}
\end{align}
where $R$ is the scalar curvature, $\nabla^j$ denotes covariant differentiation, $\mu$ is the mass-energy density, and $J_i$ the components of the momentum density. Then the dominant energy condition takes the form
\begin{equation}\label{WEC}
    \mu \geq |J|_g. 
\end{equation}

\subsection{Null expansions and apparent horizons}

Given a two dimensional surface $S$ inside $M$ one can calculate the future (+) and past (-) null expansion at each point of the surface. These are defined by
\begin{equation*}
    \theta_\pm = H_S \pm Tr_S k
\end{equation*}
where $H_S$ indicates the mean curvature of the surface, and $Tr_S k$ indicates the trace of $k$ restricted to $S$ calculated with respect to the induced metric. The null expansions measure the convergence and divergence of past and future directed null geodesics. A future or past apparent horizon is defined by  
\begin{equation*}
    \theta_{\pm}=H_S \pm Tr_S k = 0
\end{equation*}
and this is a popular way of modeling black holes in initial data sets without knowning the full development of the initial data.

\subsection{Asymptotically Flat Initial Data Sets in Spherical Symmetry}

We take our manifold to be
\begin{equation} \label{Manifold} 
    M=\mathbb{R}^3\setminus B_1 (0)=\lbrace (r, \theta, \phi): r\in[1, \infty), \theta\in (0, \pi), \phi \in (0, 2\pi) \rbrace
\end{equation}
with general spherically symmetric metric
\begin{align} \label{SphericalMEtric}
g=g_{11}(r)dr^2 + \rho^2(r) d\theta^2 + \rho^2(r) \sin^2 (\theta) d\phi^2
\end{align}
and spherically symmetric extrinsic curvature
\begin{align} \label{SphericalExtrinsicCurvature}
    k= g_{11} k_a dr^2 + k_b \rho^2 d\theta^2 + k_b \rho^2 \sin^2 (\theta) d\phi^2 
\end{align}
where $g_{11}(r)>0$, $\rho(r)>0$, and $k_a=k_a(r), k_b=k_b(r)$ are some arbitrary functions of $r$. We assume the usual fall-off conditions
\begin{align} \label{FallOffConditions} \begin{split}
    |k(r)|_g\leq Cr^{-2}, \quad &|Tr_g k(r)|\leq Cr^{-3},  \quad |(g_{11}-1)(r)|+r|{g_{11, r}(r)}|\leq Cr^{-1}, \\ & \quad |\rho(r)-r|+ r|\rho_r (r) -1 | + r^2 |\rho_{rr}(r)| \leq C \end{split} 
\end{align}
for some constant $C$ which simply say that the initial data set is asymptotically flat. It is also easy to calculate $Tr_g k$ in such coordinates and we find
\begin{equation*}
    Tr_g k = k_a + 2k_b.
\end{equation*}

The null expansions for a sphere $S_r$ of coordinate radius $r$ are given by
\begin{align*}
    \theta_{\pm}(S_r)=\theta_{\pm}(r) = 2\left( \sqrt{g^{11}} \frac{\rho_r}{\rho  } \pm k_b   \right)(r).
\end{align*}
If the boundary $\partial M =S_1$ is an outermost (future or past or both) apparent horizon, then
\begin{align} \label{OutermostCondition}
    \theta_{\pm}(r)>0, \quad r>1
\end{align}
and 
\begin{align*}
    H_{S_r} = \frac{1}{2} \left(  \theta_+ + \theta_- \right) =  2\sqrt{g^{11}}\frac{\rho_r}{\rho  } >0, \quad r>1.  
\end{align*}

\subsection{The Generalized Jang Equation in Spherical Symmetry}

For any $\phi>0$, defining 
\begin{align} \label{DefinitionOfv}
 v(r)=   \frac{  \phi \sqrt{g^{11}} f_{r}     }{ \sqrt{1+\phi^2  g^{11}  f_{r}^2   }         }
\end{align}
the generalized Jang equation takes the form
\begin{align} \label{SphericalGeneralizedJangeEq}
    \sqrt{g^{11}} v_r + 2 \left( \sqrt{g^{11}} \frac{\rho_r}{\rho} v - k_b  \right) + (v^2-1)k_a + \sqrt{g^{11}} v (1-v^2) \frac{\phi_r}{\phi}  = 0.
\end{align}
See equation (7) in \cite{BrayKhuri}. Using the fact that $Tr_g k = k_a +2 k_b$, we can also write this as 
\begin{equation} \label{SphericalGJE2}
    \sqrt{g^{11}} v_r + 2 \sqrt{g^{11}} \frac{\rho_r}{\rho} v  + v^2k_a - Tr_g k  + \sqrt{g^{11}} v (1-v^2) \frac{\phi_r}{\phi}  = 0
\end{equation}
We can calculate a few useful quantities in terms of $v$. We have
\begin{align*}
    \bar{g}_{11} = g_{11} + \phi^2 f_r^2 = g_{11} (1+ \phi^2 g^{11} f_r^2)=g_{11} \left(\frac{1}{1-v^2}\right)
\end{align*}
and
\begin{align*}
    g_{22}=\bar{g}_{22}=\rho^2, \quad  g_{33}=\bar{g}_{33}=\rho^2 \sin^2(\theta), \quad  g_{ij}=\bar{g}_{ij}=0, \quad i\neq j.
\end{align*}
so 
\begin{equation*}
    \bar{g}= g_{11} \left(\frac{1}{1-v^2}\right)dr^2 + \rho^2(r)  d\theta^2 + \rho^2(r)\sin^2(\theta) d\phi^2
\end{equation*}
and 
\begin{equation}
    |\bar{g}|=\frac{|g|}{1-v^2}.
\end{equation}
Notice that for $v$ to yield a smooth $f$, one needs 
\begin{equation*}
    -1< v(r) < 1.
\end{equation*}

\subsection{The zero divergence term in spherical symmetry}

In spherical symmetry one has $q_2=q_3=0$ and 
\begin{equation*}
    q_1= -2\sqrt{g_{11} }\frac{v}{1-v^2} \left(  \sqrt{g^{11}} \frac{\rho_r}{\rho} v -k_b  \right)
\end{equation*}
(see Appendix C of \cite{BrayKhuri}) and raising the index using $\bar{g}$ we have
\begin{equation*}
    q^1=\bar{g}^{11}q_1 =-2\sqrt{g^{11}}v\left(   \sqrt{g^{11}}\frac{\rho_r}{\rho} v -k_b \right).
\end{equation*}
So the zero divergence condition is
\begin{equation*}
    \overline{div}(\phi q ) = \frac{1}{\sqrt{|\bar{g}|}} \partial_r \left(  \sqrt{|\bar{g}|} \phi q^1     \right) = 0
\end{equation*}
which we can rewrite as 
\begin{equation} \label{RelationPhi1}
    \partial_r(\sqrt{|\bar{g}|})\phi q^1 + \sqrt{|\bar{g}|} \phi_r q^1 + \sqrt{|\bar{g}|} \phi q^1_r = 0
\end{equation}
or 
\begin{equation} \label{RelationPhi2}
    \frac{\phi_r}{\phi} = - \frac{q^1_r}{q^1} - \frac{  \partial_r(\sqrt{|\bar{g}|})  }{ \sqrt{|\bar{g}|} }.
\end{equation}
We remark that we have to be a bit careful here. To apply the argument to the Penrose inequality, we need $\phi(r)>0$ for $r>1$. Also, $|\bar{g}|>0$ automatically for $-1<v<1$. However, to obtain \eqref{RelationPhi2}, we need to divide by $q^1$. But, \eqref{RelationPhi1} can hold if $q_1 \equiv 0$. We will see how to handle this possibility later on.

We calculate:
\begin{align*}
    \frac{q^1_r}{q^1} &= \frac{  \partial_r (  \sqrt{g^{11}} )    }{    \sqrt{  g^{11} }  } + \frac{v_r}{v} + \frac{  \partial_r \left(   \sqrt{g^{11}}\frac{\rho_r}{\rho} v -k_b \right)  }{   \left(   \sqrt{g^{11}}\frac{\rho_r}{\rho} v -k_b \right)} \\
    &= \frac{  \partial_r (  \sqrt{g^{11}} )    }{    \sqrt{  g^{11} }  } + \frac{v_r}{v} + \frac{      \sqrt{g^{11}}\frac{\rho_r}{\rho}    }{   \left(   \sqrt{g^{11}}\frac{\rho_r}{\rho} v -k_b \right)}v_r +  \frac{     \partial_r \left( \sqrt{g^{11}}\frac{\rho_r}{\rho}  \right)  }{   \left(   \sqrt{g^{11}}\frac{\rho_r}{\rho} v -k_b \right)} v -  \frac{     \partial_r \left( k_b  \right)  }{   \left(   \sqrt{g^{11}}\frac{\rho_r}{\rho} v -k_b \right)} \\
    & = \left(  \frac{1}{v} + \frac{      \sqrt{g^{11}}\frac{\rho_r}{\rho}    }{   \left(   \sqrt{g^{11}}\frac{\rho_r}{\rho} v -k_b \right)} \right)v_r +  \frac{  \partial_r (  \sqrt{g^{11}} )    }{    \sqrt{  g^{11} }  } +  \frac{     \partial_r \left( \sqrt{g^{11}}\frac{\rho_r}{\rho}  \right)  }{   \left(   \sqrt{g^{11}}\frac{\rho_r}{\rho} v -k_b \right)} v -  \frac{     \partial_r \left( k_b  \right)  }{   \left(   \sqrt{g^{11}}\frac{\rho_r}{\rho} v -k_b \right)}.
\end{align*}
Also
\begin{align*}
    \frac{  \partial_r(\sqrt{|\bar{g}|})  }{ \sqrt{|\bar{g}|} } & = \frac{  \partial_r\left( 
 (1-v^2)^{-1/2}\sqrt{|{g}|}    \right)  }{ (1-v^2)^{-1/2} \sqrt{|{g}|} } = \frac{  \partial_r\left( 
 (1-v^2)^{-1/2}\sqrt{g_{11}} \rho^2 \sin(\theta)    \right)  }{ (1-v^2)^{-1/2} \sqrt{g_{11}} \rho^2 \sin(\theta)  } \\
 &= \frac{\partial_r \left(  (1-v^2)^{-1/2} \right)}{(1-v^2)^{-1/2}} + \frac{\partial_r(\sqrt{g_{11}})}{\sqrt{g_{11}}} + 2 \frac{\rho_r}{\rho} \\
 &= \frac{v}{1-v^2} v_r + \frac{\partial_r(\sqrt{g_{11}})}{\sqrt{g_{11}}} + 2 \frac{\rho_r}{\rho}.
\end{align*}
The primary terms we want to focus on are the terms containing $v_r$. Thus we can write
\begin{align} \label{PhiRelation}
    \frac{\phi_r}{\phi} = -  \left( \frac{v}{1-v^2} + \frac{1}{v} + \frac{      \sqrt{g^{11}}\frac{\rho_r}{\rho}    }{   \left(   \sqrt{g^{11}}\frac{\rho_r}{\rho} v -k_b \right)} \right)v_r - F
\end{align}
where
\begin{equation*}
    F=   \frac{     \partial_r \left( \sqrt{g^{11}}\frac{\rho_r}{\rho}  \right)  }{   \left(   \sqrt{g^{11}}\frac{\rho_r}{\rho} v -k_b \right)} v -  \frac{     \partial_r \left( k_b  \right)  }{   \left(   \sqrt{g^{11}}\frac{\rho_r}{\rho} v -k_b \right)} + 2\frac{\rho_r}{\rho}. 
\end{equation*}
where we used the fact that
\begin{equation*}
    \frac{\partial_r(\sqrt{g_{11}})}{\sqrt{g_{11}}} + \frac{\partial_r(\sqrt{g^{11}})}{\sqrt{g^{11}}} = 0.
\end{equation*}
Substituting \eqref{PhiRelation} into \eqref{SphericalGJE2} we obtain
\begin{Small}
\begin{align*}
     \sqrt{g^{11}} v_r + 2 \sqrt{g^{11}} \frac{\rho_r}{\rho} v  + v^2k_a  - Tr_g k   - \sqrt{g^{11}} v (1-v^2) \left(  \left( \frac{v}{1-v^2} + \frac{1}{v} + \frac{      \sqrt{g^{11}}\frac{\rho_r}{\rho}    }{   \left(   \sqrt{g^{11}}\frac{\rho_r}{\rho} v -k_b \right)} \right)v_r + F  \right) =0
\end{align*}
\end{Small}
or, after a bit of algebra, 
\begin{align} \label{DegenerateODE}
    \sqrt{g^{11}} \left( (v^2-1)  \frac{      \sqrt{g^{11}}\frac{\rho_r}{\rho} v    }{   \left(   \sqrt{g^{11}}\frac{\rho_r}{\rho} v -k_b \right)}  \right)v_r  + 2 \sqrt{g^{11}} \frac{\rho_r}{\rho} v  + v^2k_a  - Tr_g k - \sqrt{g^{11}} v (1-v^2)F =0.
\end{align}
The basic problem with this equation is that if $(v^2-1)<0$, as is needed for smooth solutions, the coefficient of $v_r$ has the wrong sign.

\subsection{Construction of the spherically symmetric initial data set}

We want our spherically symmetric initial data set to be asymptotically flat,  satisfy the dominant energy condition, and have the boundary consist of an outermost apparent horizon. To do this, we use methods similar to those of \cite{JaraczWEC}.

To simplify things, we let $\rho(r)=r$, so our metric has the form
\begin{align} \label{SphericalMetric2}
    g=h(r)dr^2 + r^2 d\theta^2 + r^2 \sin^2 (\theta) d\phi^2
\end{align}
where we let $g_{11}=h$ and 
\begin{align} \label{SphericalExtrinsicCurvature2}
     k= h k_a dr^2 + k_b r^2 d\theta^2 + k_b r^2 \sin^2 (\theta) d\phi^2. 
\end{align}
In such a metric we can explicitly calculate $J_i$. We find that $J_2=J_3=0$ and 
\begin{align*}
    J_1 &= \frac{1}{8 \pi} \left(  \partial_r(k_a) + 2\frac{k_a - k_b}{r} - \partial_r (Tr_g k )  \right) \\
    &=\frac{1}{8 \pi} \left(  \partial_r(k_a) + 2\frac{k_a - k_b}{r} - \partial_r (k_a + 2k_b )  \right) \\ 
    &= \frac{1}{4\pi} \left( \frac{k_a - k_b}{r} - \partial_r(k_b)    \right).
\end{align*}

We want $k_b$ to be compactly supported. This is extremely useful because when $k_b=0$ and $\rho(r)=r$, the equation \eqref{DegenerateODE} simplifies greatly. Let $\Phi(r)$ be a smooth cut-off function satisfying:
\begin{align*}
    \Phi(r) = \begin{dcases}  
    1 &: \quad 1\leq r \leq 2 \\
    \text{smooth, decreasing} &: \quad 2<r<3 \\
    0 &: \quad r\geq 3 \end{dcases}.
\end{align*}
Define 
\begin{equation} \label{Defka}
    k_a \coloneqq \frac{1}{6} \left(  \frac{5 \sin(r)}{r^{5.1}} - \frac{\cos(r)}{r^{4.1}}    \right)
\end{equation}
and \begin{equation} \label{defkb}
    k_b \coloneqq -\Phi(r). 
\end{equation}
From this point forward, when we write $k_a, k_b$ we mean the functions given by these definitions. Now, this choice of $k_a$ might seem extremely perplexing at first. However, it is chosen so that 
\begin{equation}
    \int (- r^{-0.9} k_a(r)) dr = \frac{\sin(r)}{6 r^5} + C
\end{equation}
as can be easily checked, which becomes extremely useful in the calculations \eqref{Calculation1} and \eqref{Calculation2}.
Also
\begin{equation} \label{boundka}
    |k_a| \leq \frac{1}{r^{4.1}}
\end{equation}

Notice, with these definitions we have
\begin{align*}
    Tr_g k = k_a + 2k_b 
\end{align*}
and
\begin{align*}
    Tr_g k = k_a  \quad r\geq 3.
\end{align*}
With the above $k_a, k_b$ define
\begin{equation*}
    U(r) = \frac{k_a - k_b}{r} - \partial_r(k_b) 
\end{equation*}
and notice $|U(r)|<1/r^5$ for $r\geq 3$. Then
\begin{equation*}
    |J|_g^2 = g^{ij}J_i J_j = g^{11}J_1J_1 = \frac{1}{h} \frac{1}{16 \pi^2} U^2 
\end{equation*}
and so 
\begin{equation*}
    |J|_g = \frac{1}{4\pi}\frac{1}{\sqrt{h}} |U|.
\end{equation*}
Next, take some fixed smooth function $V(r)>0$ defined for  $r\geq 1$ with $V(r)\geq |U(r)|$ and $V(r)=1/r^4$ for $r\geq 4$ which is possible by \eqref{boundka}. Then
\begin{equation*}
    \frac{1}{4\pi} \frac{1}{\sqrt{h}} V \geq |J|_g
\end{equation*}
Next we need to find an $h$ which will satisfy the dominant energy condition. In spherical symmetry we have
\begin{align*}
    16\pi \mu &= R + (Tr_g k)^2 - |k|_g^2 = R + (k_a + 2 k_b)^2 - ( k_a^2 + 2k_b^2  ) \\ 
    & = R + 4k_ak_b + 2k_b^2.
\end{align*}
Notice, since $k_b$ is supported on the interval $[1, 3]$ so is $4k_a k_b +2k_b^2$. Now, take some fixed smooth function $W(r)$ defined for  $r\geq 1$ with $W(r)\geq |4k_a k_b + 2k_b^2|$ and compactly supported on $[1, 4)$. We need $16\pi \mu \geq 16\pi |J|_g$. Therefore we can take:
\begin{equation*}
    R(r) = W(r) + \frac{4V(r)}{\sqrt{h(r)}}
\end{equation*}
since then 
\begin{equation*}
    16\pi \mu = W(r) + \frac{4V(r)}{\sqrt{h(r)}} + 4k_a k_b + 2k_b^2 \geq \frac{4V(r)}{\sqrt{h(r)}} \geq  \frac{4|U(r)|}{\sqrt{h(r)}} = 16\pi |J|_g.
\end{equation*}
Now, define 
\begin{equation*}
    k_\epsilon = \epsilon k
\end{equation*}
so that $k_{a\epsilon}=\epsilon k_a$ and $k_{b\epsilon}=\epsilon k_b$. In this case
\begin{align*}
    Tr_g k_\epsilon = \epsilon Tr_g k, \quad |J_\epsilon|_g = \epsilon |J|_g, \quad 4k_{a\epsilon} k_{b\epsilon} +2k_{b\epsilon}^2=\epsilon^2 (4k_a k_b +2k_b^2) 
\end{align*}
and so if we had a metric $g_\epsilon$ with the prescribed scalar curvature
\begin{equation} \label{prescribedScalarCurvature}
     R_\epsilon (r) = \epsilon^2W(r) + \frac{4\epsilon V(r)}{\sqrt{h(r)}}
\end{equation}
then $(M, g_\epsilon, k_\epsilon)$ would satisfy the dominant energy condition.

For a metric of the form \eqref{SphericalMetric2} the scalar curvature is given by
\begin{equation*}
    R(r)= \frac{2h'(r)}{rh^2(r)} - \frac{2}{r^2 h(r)} + \frac{2}{r^2}
\end{equation*} 
(see equation (2.10) in \cite{JaraczWEC} and the equation before (20) in \cite{BrayKhuri}) which we can rearrange to obtain
\begin{equation} \label{ODE}
    h'=\frac{h}{r} - \frac{h^2}{r} + \frac{1}{2}Rrh^2.
\end{equation}
Substituting \eqref{prescribedScalarCurvature} we obtain
\begin{equation} \label{ODE2}
    h'=\frac{h}{r}-\frac{h^2}{r} + \frac{1}{2}\epsilon^2 rW h^2 + 2\epsilon r V h^{3/2}.
\end{equation}
It turns out that for sufficiently small $\epsilon>0$ this ordinary differential equation can be solved for any initial condition $h(1)>0$.

\begin{proposition} \label{PropositionODEExistence}
There exists some $\epsilon>0$ such that \eqref{ODE2} has a smooth solution $h(r)$ for any $h(1)>0$. Moreover
\begin{align} \label{AsymptoticFlatnessCondition}
     |h(r)-1|+r|{h_{r}(r)}|\leq Cr^{-1}
\end{align}
for some constant $C$ depending on $h(1)$. 
\end{proposition}

The proof is an application of some basic methods of ordinary differential equations. However, written out in full detail it becomes quite lengthy. Thus, in order to not interrupt the flow of the paper, we relegate it to Appendix \ref{AppendixA}. To finish constructing our initial data set, we must now pick the correct initial condition for $h(1)$ at the boundary, and then take the region exterior to the outermost apparent horizon.

\begin{proposition} \label{PropInitialDataSet1}
    Let $(M, g_\epsilon, k_\epsilon)$ be an initial data set with $M$ given by \eqref{Manifold}, $k_\epsilon=\epsilon k$ where $k$ is given by \eqref{SphericalExtrinsicCurvature2} with $k_a$ and $k_b$ given by \eqref{Defka} and \eqref{defkb} respectively, and $g_\epsilon$ given by \eqref{SphericalMetric2} with $h(r)$ being the solution given in Proposition \ref{PropositionODEExistence} with $h(1)=(1/\epsilon)^2$. Then $(M, g_\epsilon, k_\epsilon)$ is an asymptotically flat initial data set satisfying the dominant energy condition, containing no compact past apparent horizons, and with $\partial M$ being a compact future apparent horizon.   
\end{proposition}

\begin{proof}
We have already proven all the statements, by construction, except for the last two. Due to asymptotic flatness, there always exists an outermost future and an outermost past apparent horizon. These are unique, and possibly empty, and with possibly several components \cite{AnderssonMetzger}. However, due to the uniqueness and spherical symmetry, these must consist of some coordinate spheres $S_r$.   Notice, for \eqref{SphericalMetric2}, the mean curvature of a coordinate sphere is
\begin{equation*}
    H_{S_r} =   \frac{2}{r \sqrt{h(r)}} > 0
\end{equation*}
since $h(r)>0$. Also
\begin{equation*}
    \theta_{\pm}(S_r) = \theta_{\pm}(r) =  \frac{2}{r \sqrt{h(r)}} \pm 2k_{b\epsilon}(r)= \frac{2}{r \sqrt{h(r)}} \pm 2\epsilon k_{b}(r) 
\end{equation*}
but since $k_b\leq 0$ then $\theta_-(r)>0$ and so there is no outermost compact past horizon, and hence no compact past horizons whatsoever. 

Also, with our choice of $h(1)$ we have
\begin{equation*}
    \theta_+(1)=\frac{2}{\sqrt{h(1)}} - 2\epsilon \Phi(1) = \frac{2}{\sqrt{(1/\epsilon)^2}} - 2 \epsilon = 0
\end{equation*}
so indeed $S_1=\partial M$ is a compact future apparent horizon.
\end{proof}

Now, $S_1$ might not be an outermost future apparent horizon. However, as mentioned, there is always an outermost future apparent horizon, and due to the spherical symmetry is some sphere of coordinate radius $r_0$. Therefore if we take the region exterior to this horizon, we have the following:

\begin{proposition} \label{PropositionInitialDataSet}
    Let $S_{r_0}$ denote the outermost apparent horizon of $(M, g_\epsilon, k_\epsilon)$ of Proposition \ref{PropInitialDataSet1}. Let $M_\epsilon=\lbrace r\geq r_0\rbrace \subset M$. Then $(M_\epsilon, g_\epsilon, k_\epsilon)$ is an asymptotically flat initial data set satisfying the dominant energy condition with boundary consisting of an outermost compact future apparent horizon, and containing no other compact apparent horizons.
\end{proposition}

\subsection{Appropriate Asymptotics for $v(r)$} In order to apply the Riemannian Penrose inequality in the Jang surface to conclude that the Penrose inequality holds for the original data set, the ADM energies of $(M, g, k)$ and $(M, \bar{g})$ must be the same. This means that $v(r)$ must have certain asymptotics at infinity. 

If we write $h(r)=1+\psi(r)$, after a tedious calculation, the ADM energy of the metric \eqref{SphericalMetric2} is given by
\begin{equation*}
    E_{ADM} = \lim_{r \rightarrow \infty} \frac{r}{2} \frac{\psi(r)}{\sqrt{h(r)}}.
\end{equation*}
This formula can also be verified by taking the limit of the Hawking mass which is asymptotic to the ADM energy in the case of asymptotic flatness, giving
\begin{align*}
    \lim_{r \rightarrow \infty} M_H(S_r) &= \lim_{r \rightarrow \infty} \sqrt{\frac{|S_r|}{16\pi}} \left( 1 -  \frac{1}{16 \pi} \int_{S_r} H^2 dS_r   \right) 
     =\lim_{r \rightarrow \infty} \frac{r}{2} \left( 1 - \frac{1}{h(r)}  \right) \\ 
     &= \lim_{r \rightarrow \infty} \frac{r}{2} \left( 1 - \frac{1}{h(r)}  \right) = \lim_{r \rightarrow \infty} \frac{r}{2} \frac{\psi(r)}{h(r)}.  
\end{align*}
In the asymptotically flat case where $\lim_{r \rightarrow \infty} h(r)=1$ both of these formulas yield the same limit, as they should. 

Therefore, if we look at the metric of the Jang surface, where $\bar{h}=h+ \phi^2 f_r^2=1+\psi + \phi^2 fr^2$, we need 
\begin{equation} \label{A1}
    \lim_{r \rightarrow \infty} \phi^2(r) f_r^2(r) = 0
\end{equation}
to ensure asymptotic flatness. Moreover
\begin{align*}
    \bar{E}_{ADM} &= \lim_{r \rightarrow \infty}  \frac{r}{2} \frac{\psi(r)+\phi^2(r) f_r^2(r)}{\sqrt{1+\psi(r)+ \phi^2(r)f_r^2(r)}} = \lim_{r \rightarrow \infty} \frac{r}{2}(\psi(r) + \phi^2(r) f_r^2(r)  ) \\ &= E_{ADM} + \lim_{r\rightarrow \infty} \frac{r \phi^2(r) f_r^2(r)}{2}
\end{align*}
and so to have the ADM energies match up we need
\begin{equation*}
    {\phi^2(r) f_r^2(r)} \leq \frac{C}{r^{1+2\varepsilon}}
\end{equation*}
for some $\varepsilon>0$, or using \eqref{DefinitionOfv} and \eqref{A1} we need
\begin{equation} \label{AppropriateAsymptoticsv}
    |v(r)| \leq \frac{C}{r^{ 1/2 + \varepsilon}}.
\end{equation}

\subsection{Proof of Theorem \ref{MainTheorem}}

\begin{proof}

We will take the initial data set $(M_\epsilon, g_\epsilon, k_\epsilon)$ of Proposition \ref{PropositionInitialDataSet} and show that there are no smooth solutions $f=f(r)$ and  $ \phi=\phi(r)>0$ for $r\in (r_0, \infty)$ to the system \eqref{CoupledSystem} having the asymptotics \eqref{AppropriateAsymptoticsv} for $v(r)$. The proof is by contradiction. 

Suppose smooth solutions $ f(r)$ and $ \phi(r)$ with $\phi(r)>0$ solving \eqref{CoupledSystem} with $v(r)$ satisfying \eqref{AppropriateAsymptoticsv} exist for $(M_\epsilon, g_\epsilon, k_\epsilon)$. Let us consider the solutions on the interval $[4, \infty)$ since  
\begin{equation*}
    k_{b \epsilon}(r) = 0, \quad Tr_{g_{\epsilon}} k_\epsilon (r) = \epsilon k_a(r) \quad \text{for} \quad  r\geq 4
\end{equation*}
which makes the analysis considerably easier.

First, notice that $v(r) \not\equiv 0$ on $[4, \infty)$. For substituting $v(r)\equiv 0$ into \eqref{SphericalGJE2} yields
\begin{equation*}
    \epsilon k_a = Tr_{g_{\epsilon}} k_\epsilon \equiv 0, \quad r\geq 4
\end{equation*}
which is false by construction. Thus, we can take some point $s_1\in[4, \infty)$ where $v(s_1)\neq 0$. Without loss of generality, we can assume $v(s_1)>0$. Now, let $I=[s_1, s_*) \subset [s_1, \infty)$ be the maximal interval on which $v(r)>0$.  Suppose $s_* < \infty$ so that $v(s_*)=0$. In that case, \eqref{PhiRelation} must hold for $r\in [s, s_*)$. Moreover, since $r\geq 4$, the equations simplify considerably. 

We have 
\begin{align} \label{SimpleF}
    F= \frac{\partial_r ( h^{-1/2} r^{-1}) }{h^{-1/2}r^{-1}} + \frac{2}{r}
\end{align}
and so 
\begin{equation*}
    \frac{\phi_r}{\phi} = -\frac{2 v_r}{ v} - \frac{v v_r}{1-v^2} - \frac{\partial_r ( h^{-1/2} r^{-1}) }{h^{-1/2}r^{-1}} - \frac{2}{r} 
\end{equation*}
and therefore integrating 
\begin{align*}
    \ln(\phi) & =-2\ln(v) + \frac{1}{2}\ln(1-v^2) - \ln(h^{-1/2}r^{-1}) - 2 \ln(r) + C_1 \\
    & = \ln\left(  \frac{r \sqrt{h} \sqrt{1-v^2}}{r^2 v^2} \right) + C_1
\end{align*}
and so
\begin{equation*}
    \phi(r)= C_2 \frac{  \sqrt{h(r)} \sqrt{1-v^2(r)}}{r v^2(r)}, \quad r\in[s_1, s_*)
\end{equation*}
for some $C_2>0$. But since 
\begin{equation*}
    \lim_{r \rightarrow s_*^-} v(r) = v(s_*)=0
\end{equation*}
then
\begin{equation*}
    \lim_{r \rightarrow s_*^-} \phi(r) = \infty
\end{equation*}
contradicting the assumed smoothness of $\phi(r)$. The same argument works if we assume $v(s_1)<0$.

Therefore, we can assume we are on some interval $[s_1, \infty)\subset[4, \infty)$ where $v(r)\neq 0$. First we assume $v(r)>0$ on $[s_1, \infty)$. In that case, $q^1(r) \neq 0$ and $v(r)$ must satisfy \eqref{DegenerateODE} on $[s_1, \infty)$. In this case, \eqref{DegenerateODE} becomes
\begin{align*}
    \frac{1}{\sqrt{h}} (v^2-1)v_r + \frac{2}{r \sqrt{h}} v + \epsilon k_a v^2 - \epsilon k_a - \frac{1}{\sqrt{h}} v(1-v^2) \left( \frac{\partial_r(h^{-1/2}r^{-1})}{h^{-1/2} r^{-1}} + \frac{2}{r}    \right) = 0
\end{align*}
which we can further simplify to get
\begin{align*}
    \frac{1}{\sqrt{h}} (v^2-1)v_r + \frac{2}{r \sqrt{h}} v + \epsilon k_a v^2 - \epsilon k_a - \frac{1}{\sqrt{h}} v(1-v^2) \left( \frac{1}{r} - \frac{h_r}{2h}    \right) = 0.
\end{align*}
We can then put the equation in the form
\begin{align}
    \frac{1}{\sqrt{h}}v_r - \left( \frac{2}{r(1-v^2)} - \frac{1}{r} + \frac{h_r}{2h}   \right)\frac{v}{\sqrt{h}} + \epsilon k_a = 0.
\end{align}
Next, to make things easier, we let
\begin{equation*}
    \tilde{v}(r)=\frac{v(r)}{\sqrt{h(r)}}
\end{equation*}
in which case 
\begin{equation*}
    v_r = \sqrt{h} \tilde{v}_r + \frac{1}{2}h^{-1/2}h_r \tilde{v} 
\end{equation*}
which upon substituting gives
\begin{equation}
    \tilde{v}_r - \left(  \frac{2}{r(1-h\tilde{v}^2)}  - \frac{1}{r}\right) \tilde{v} + \epsilon k_a = 0.
\end{equation}

Now, if $\lim_{r \rightarrow \infty} v(r) \neq 0$ then the solution does not have appropriate asymptotics for application to the Penrose inequality. Hence, we can assume $\lim_{r \rightarrow \infty} v(r) = 0$ and since $\lim_{r \rightarrow \infty} h(r)=1$ we also have $\lim_{r \rightarrow \infty} \tilde{v}(r) = 0$. Thus, we can take some interval $[s_2, \infty) \subset [s_1, \infty)$ such that 
\begin{equation} \label{InequalityODE}
     \frac{2}{r(1-h\tilde{v}^2)}  - \frac{1}{r} > \frac{0.9}{r} \quad \text{for} \quad r\geq s_2.
\end{equation}
Now take some $s_3>s_2$ such that
\begin{equation*}
    \frac{\sin(s_3)}{s_3^5}<0
\end{equation*} and consider the initial value problem
\begin{align*}
    \underline{w}_r - \frac{0.9}{r} \underline{w} +\epsilon k_a &= 0 \\
    \underline{w}(s_3) &= \frac{1}{2} \tilde{v}(s_3).
\end{align*}
The solution can be explicitly calculated using the method of integrating factors to be 
\begin{align} \begin{split} \label{Calculation1}
    \underline{w}(r)&= r^{0.9} \left[ \frac{\tilde{v}(s_3)}{2s_3^{0.9}} + \int_{s_3}^r (-\epsilon s^{-0.9} k_a(s)) ds   \right] \\
    &=  r^{0.9} \left[ \frac{\tilde{v}(s_3)}{2s_3^{0.9}} + \frac{\epsilon \sin(r)}{6 r^5} -  \frac{ \epsilon \sin(s_3)}{ 6 s_3^5}    \right] \\
    &= r^{0.9} \left[ P + \frac{\epsilon \sin(r)}{6 r^5}  \right] \end{split}
\end{align}
for $r\geq s_3$, where, since $v(s_3)>0$ and $\sin(s_3)/s_3^5 <0$, we have $P>0$ is a positive constant. Notice, this calculation is precisely why we defined $k_a$ in \eqref{Defka} the way we did.
Also notice we have
\begin{equation*}
    \lim_{r \rightarrow \infty} \underline{w}(r) = \infty
\end{equation*}
because $P>0$.

We claim that $\tilde{v}(r) > \underline{w}(r)$ for $r\geq s_3$. Notice, $\tilde{v}(s_3) > \underline{w}(s_3)$. Now let $\tilde{s}> s_3$ be the smallest $r>s_3$ where $\underline{w}(\tilde{s})=\tilde{v}(\tilde{s})= \beta$. At such a point we must have
\begin{equation*}
    \tilde{v}_r ( \tilde{s}) \leq \underline{w}_r(\tilde{s})
\end{equation*}
Notice, since we are assuming $v(r)>0$ on $[s_1, \infty)$ then necessarily $\beta>0$. Therefore, the following inequality holds:
\begin{align*}
    \tilde{v}_r (\tilde{s}) = \left(\frac{2}{\tilde{s}(1-h({\tilde{s}})\tilde{v}^2(\tilde{s}))}  - \frac{1}{\tilde{s}} \right) \beta - \epsilon k_a(\tilde{s}) > \frac{0.9}{\tilde{s}} \beta - \epsilon k_a(\tilde{s}) = \underline{w}_r (\tilde{s})
\end{align*}
due to \eqref{InequalityODE} which is a contradiction. Therefore, $\tilde{v}(r) > \underline{w}(r)$ for $r\geq s_3$ and so $v(r) > \sqrt{h(r)}\underline{w}(r)$ for $r\geq s_3$. Therefore
\begin{equation*}
    \lim_{r\rightarrow \infty} v(r) = \infty
\end{equation*}
contradicting the assumption $\lim_{r\rightarrow \infty} v(r) = 0$. Thus, assuming $v(r)>0$ on $[s_1, \infty)$ yields a contradiction to the necessary asymptotics of \eqref{AppropriateAsymptoticsv}.

So finally, we are left with the case $v(r)<0$ on $[s_1, \infty)$. By the same arguments, we can assume we are on the interval $[s_2, \infty)$ where \eqref{InequalityODE} holds. Now take some $s_4>s_2$ such that
\begin{equation*}
    \frac{\sin(s_4)}{s_4^5}>0
\end{equation*} and consider the initial value problem
\begin{align*}
    \overline{w}_r - \frac{0.9}{r} \overline{w} +\epsilon k_a &= 0 \\
    \overline{w}(s_4) &= \frac{1}{2} \tilde{v}(s_4).
\end{align*}
Therefore
\begin{align} \begin{split} \label{Calculation2}
    \overline{w}(r)&= r^{0.9} \left[ \frac{\tilde{v}(s_4)}{2s_4^{0.9}} + \int_{s_4}^r (-\epsilon s^{-0.9} k_a(s)) ds   \right] \\
    &=  r^{0.9} \left[ \frac{\tilde{v}(s_4)}{2s_4^{0.9}} + \frac{\epsilon \sin(r)}{6 r^5} -  \frac{ \epsilon \sin(s_4)}{ 6 s_4^5}    \right] \\
    &= r^{0.9} \left[ N + \frac{\epsilon \sin(r)}{6 r^5}  \right] \end{split}
\end{align}
for $r\geq s_4$, where, since $v(s_4)<0$ and $\sin(s_4)/s_4^5 >0$, we have $N<0$ is a negative constant. Therefore we have
\begin{equation*}
    \lim_{r \rightarrow \infty} \overline{w}(r) = -\infty
\end{equation*}
because $N<0$.

We claim that $\tilde{v}(r) < \overline{w}(r)$ for $r\geq s_4$. Notice, $\tilde{v}(s_4) < \overline{w}(s_4)$. Now let $\tilde{s}> s_4$ be the smallest $r>s_4$ where $\overline{w}(\tilde{s})=\tilde{v}(\tilde{s})= \gamma$. At such a point we must have
\begin{equation*}
    \tilde{v}_r ( \tilde{s}) \geq \overline{w}_r(\tilde{s})
\end{equation*}
Notice, since we are assuming $v(r)<0$ on $[s_1, \infty)$ then necessarily $\gamma<0$. Therefore, the following inequality holds:
\begin{align*}
    \tilde{v}_r (\tilde{s}) = \left(\frac{2}{\tilde{s}(1-h({\tilde{s}})\tilde{v}^2(\tilde{s}))}  - \frac{1}{\tilde{s}} \right) \gamma - \epsilon k_a(\tilde{s}) < \frac{0.9}{\tilde{s}} \gamma - \epsilon k_a(\tilde{s}) = \underline{w}_r (\tilde{s})
\end{align*}
since multiplying \eqref{InequalityODE} by $\gamma<0$ reverses the inequality sign. This is again a contradiction and therefore, $\tilde{v}(r) < \overline{w}(r)$ for $r\geq s_4$ and so $v(r) < \sqrt{h(r)}\overline{w}(r)$ for $r\geq s_4$. Therefore
\begin{equation*}
    \lim_{r\rightarrow \infty} v(r) = -\infty
\end{equation*}
contradicting the assumption $\lim_{r\rightarrow \infty} v(r) = 0$. Thus, assuming $v(r)<0$ on $[s_1, \infty)$ yields a contradiction to the necessary asymptotics of \eqref{AppropriateAsymptoticsv}.

Therefore, all the possibilities lead to contradictions, hence we conclude there are no smooth $v(r)$ and $\phi(r)$ with the appropriate asymptotics (and hence no smooth $f(r)$ and $\phi(r)$ with the appropriate asymptotics) for application to the Penrose inequality solving \eqref{CoupledSystem} for the initial data set $(M_\epsilon, g_\epsilon, k_\epsilon)$ of Proposition \eqref{PropositionInitialDataSet}. \end{proof}

\appendix

\section{Proof of Proposition \ref{PropositionODEExistence}}\label{AppendixA}

\begin{proof}
    Since $W(r)\geq 0$ is compactly supported and $V(r)>0$ is smooth with $V(r)=1/r^4$ for $r\geq 4$, there exists some constant $C_1>0$ such that 
    \begin{equation*}
        W(r) \leq \frac{C_1}{r^4}, \quad V(r) \leq \frac{C_1}{r^4}
    \end{equation*}
for $r\geq 1$.
Next, choose an $\epsilon$ so small that
\begin{equation*}
    \frac{1}{2}\epsilon^2C_1 + 2\epsilon C_1 < \frac{1}{10}
\end{equation*}
Next, let 
\begin{equation*}
    B_1 =1 + \max\lbrace h(1), 10 \rbrace. 
\end{equation*}
We claim that this constant acts as an upper barrier for the solution. Since $h(1)>0$, \eqref{ODE2} has a smooth solution on some maximal interval $[1, r^*)$. Let $s\in[1, r^*)$ be the smallest value of $r$ at which $h(r)=B_1$. Then, since the solution starts out smaller than $B_1$, we must have $h'(s)\geq 0$. However at $r=s$ we have
\begin{align*}
    h'(s)&=\frac{B_1}{r}-\frac{B_1^2}{r} + \frac{1}{2}\epsilon^2 rW B_1^2 + 2\epsilon r V B_1^{3/2} \\
    & \leq \frac{B_1}{r}-\frac{B_1^2}{r} + \frac{1}{2}\epsilon^2 r \left( \frac{C_1}{r^4}  \right) B_1^2 + 2\epsilon r \left(  
\frac{C_1}{r^4} \right) B_1^{3/2} \\
 &\leq \frac{B_1}{r}-\frac{B_1^2}{r} + \frac{1}{10} \frac{1}{r^3} B_1^2 + \frac{1}{10} \frac{1}{r^3} B_1^{3/2} \\
 & \leq \frac{B_1}{r}-\frac{B_1^2}{r} + \frac{1}{10} \frac{1}{r} B_1^2 + \frac{1}{10} \frac{1}{r} B_1^2 \\
 & = \frac{B_1}{r} \left( 1 - \frac{8B_1}{10}   \right) < 0
\end{align*}
yielding a contradiction. Hence, $h(r)<B_1$ for all $r \in [1, r^*)$. 

Similarly, we can construct a positive lower barrier. Let
\begin{equation*}
    B_0 = \frac{1}{2} \min \lbrace h(1), 1/10 \rbrace>0.
\end{equation*}
 Let $s\in[1, r^*)$ be the smallest value of $r$ at which $h(r)=B_0$. Then, since the solution starts out larger than $B_0$, we must have $h'(s)\leq 0$. However at $r=s$ we have
 \begin{align*}
     h'(s) &= \frac{B_0}{r}-\frac{B_0^2}{r} + \frac{1}{2}\epsilon^2 rW B_0^2 + 2\epsilon r V B_0^{3/2} \\
      & \geq \frac{B_0}{r}-\frac{B_0^2}{r} = \frac{B_0}{r}(1-B_0)>0
 \end{align*}
 yielding a contradiction. Hence $h(r)>B_0$ for all $r\in [1, r^*)$. 

 Thus we have $0<B_0< h(r) < B_1$ for all $r\in [1, r^*)$ which implies that $r^*=\infty$. Since $h(r)>0$, we can iteratively take as many derivatives of $h$ as we want and we conclude they are all continuous, so $h(r)$ is smooth.

 Next, we need to obtain the desired asymptotics. Notice, if $h(r_0)=1$ for some $r_0$, then $h(r)> 1$ for all $r>r_0$ since $h=1$ implies $h'>0$. Let us suppose then that we are on some interval $[r_0^*, \infty)$ with $h(r)> 1$. Consider \eqref{ODE2} for $r\geq 4$. Then the differential equation simplifies to
 \begin{align} \label{ODE4}
      h'=\frac{h}{r}-\frac{h^2}{r}  + \frac{2\epsilon}{r^3} h^{3/2}
 \end{align}
We can assume $r_0^* \geq 4$.

On this interval, we can write $h(r)=1+\psi(r)$ with $\psi>0$. Then \eqref{ODE4} can be written as
\begin{align} \label{ODE6}
    \psi'=-\frac{\psi}{r} - \frac{\psi^2}{r} + \frac{2\epsilon}{r^3}(1+\psi)^{3/2}.
\end{align}
Let $\psi(r_0^*)=\mathcal{B}>0$. Consider the initial value problem
\begin{align*}
    &w'=-\frac{w}{r} + \frac{2\epsilon B_1^{3/2}}{r^3} \\
    &w(r_0^*)=2 \mathcal{B}
\end{align*}
where $B_1$ is the upper bound for $h$ obtained earlier. We claim that $w(r)>\psi(r)$  for all $r\in [r_0^*, \infty)$. As before, let $s>r_0^*$ be the smallest value of $s$ where $w(s)=\psi(s)=B$. Since initially $w(r)$ is larger, at $s$ we must have $w'(s)\leq \psi'(s)$. But 
\begin{equation*}
    \psi'(s)=-\frac{B}{s} - \frac{B^2}{s} + \frac{2\epsilon}{s^3}(1+B)^{3/2} < -\frac{B}{s} + \frac{2\epsilon B_1^{3/2}}{s^3} = w'(s)
\end{equation*}
since $1+B=h(s)<B_1$ and so $w(r)>\psi(r)$ for $r\in [r_0^*, \infty)$. 

Now, the solution for $w(r)$ can be written down explicitly using the method of integrating factors as
\begin{align*}
    w= \frac{1}{r} \left[ 2\mathcal{B} + \int_{r_0^*}^r \frac{2\epsilon B_1^{3/2}}{s^2} ds   \right] \leq \frac{C_2}{r}
\end{align*}
which gives 
\begin{equation} \label{inequality1}
    0<\psi(r)\leq \frac{C_2}{r}
\end{equation}
for $r\geq r_0^*$.

Otherwise, we have $h(r)<1$ for all $r\in [1, \infty)$. In that case we again write $h(r)=1+\psi(r)$ with $-1< \psi(r)<0$. Let $\psi(1)=\mathcal{C}$. Consider the initial value problem
\begin{align} \label{ODE5} 
    u' = -\frac{u}{r} - \frac{u^2}{r} \\
    u(1)=\frac{1}{2} (\mathcal{C}-1).
\end{align}
 Notice, since $-1<\mathcal{C}<0$, we have $u(1)<\psi(1)$ and $-1<u(1)<0$. We claim $\psi(r)>u(r)$ for all $r\geq 1$. Again, at the first value $s$ where $u(s)=\psi(s)=B$ we'd have $u'(s)\geq \psi'(s)$. But at such a point we have
 \begin{equation*}
     u'(s)= -\frac{B}{s} - \frac{B^2}{s} < -\frac{B}{s} - \frac{B^2}{s} + \frac{1}{2}\epsilon^2 sW(s) (1+B)^2 + 2\epsilon s V(s) (1+B)^{3/2} = \psi'(s)
 \end{equation*}
since $V(r)>0$, giving a contradiction. Thus, $\psi(r)>u(r)$ for $r\geq 1$. The differential equation \eqref{ODE5} is separable. It is easy to see that if $-1<u(1)<0$ then $-1<u(r)<0$ in which case the solution can be found explicitly to be
\begin{equation*}
    u(r) = -\frac{C_3}{C_3 + r} \geq -\frac{C_3}{r}
\end{equation*}
for some $C_3>0$. Therefore
\begin{equation} \label{inequality2}
    -\frac{C_3}{r} \leq u(r) < \psi(r) <0.
\end{equation}
Putting together \eqref{inequality1} and \eqref{inequality2} gives
\begin{align} \label{inequality3}
    |\psi(r)|=|h(r)-1| \leq \frac{C_4}{r}.
\end{align}
Since $\psi'=h'$, substituting \eqref{inequality3} into \eqref{ODE6}, we get 
\begin{equation} \label{inequality4}
    |h'(r)|\leq \frac{C_5}{r^2}
\end{equation}
for $r\geq 4$. Putting together \eqref{inequality3} and \eqref{inequality4} yields \eqref{AsymptoticFlatnessCondition} for some constant $C$.
\end{proof}

\bibliographystyle{model1-num-names}
\bibliography{ref_new}

\begin{bibdiv}
\begin{biblist}

\bib{AnderssonMetzger}{article}{
      author={Andersson, L.},
      author={Metzger, J.},
       title={The area of horizons and the trapped region},
        date={2009},
     journal={Communications in Mathematical Physics},
      volume={290},
      number={3},
       pages={941\ndash 972},
}

\bib{Bartnik}{article}{
      author={Bartnik, R.},
       title={The mass of an asymptotically flat manifold},
        date={1986},
     journal={Commun. Pure Appl. Math.},
      volume={39},
       pages={661\ndash 693},
}

\bib{Bray}{article}{
      author={Bray, H.},
       title={Proof of the {Riemmanian Penrose inequality} using the positive
  mass theorem},
        date={2001},
     journal={J. Differential Geom.},
      volume={59},
      number={2},
       pages={177\ndash 267 arXiv: math/9911173},
}

\bib{BrayKhuri}{article}{
      author={Bray, H.},
      author={Khuri, M.},
       title={A {J}ang equation approach to the {P}enrose inequality},
        date={2010},
     journal={Discrete Contin. Dyn. Syst.},
      volume={27},
       pages={741\ndash 766. arXiv: 0910.4785},
}

\bib{BrayKhuriPDE}{article}{
      author={Bray, H.L.},
      author={Khuri, M.A.},
       title={P.d.e.'s which imply the penrose conjecture},
        date={2011},
     journal={Asian J. Math.},
      volume={15},
      number={4},
       pages={559\ndash 612},
}

\bib{ChruscielADM}{book}{
      author={Chru\'{s}ciel, P.T.},
      editor={Peter G.~Bergmann, Venzo~Sabbata},
       title={Boundary conditions at spatial infinity from a hamiltonian point
  of view},
        date={1986},
}

\bib{ADM}{article}{
      author={Desser{,}, R. Arnowitt{,}~S.},
      author={Misner, C.},
       title={Coordinate invariance and energy expressions in {G}eneral
  {R}elativity},
        date={1961},
     journal={Phys. Rev.},
      volume={122},
       pages={997\ndash 1006},
}

\bib{KhuriDisconzi}{article}{
      author={Disconzi, M.},
      author={Khuri, M.A.},
       title={On the penrose inequality for charged black holes},
        date={2012},
     journal={Class. Quantum Grav.},
      volume={29},
       pages={245019},
}

\bib{HanKhuri}{article}{
      author={Han, Q.},
      author={Khuri, M.A.},
       title={Existence and blow-up behavior for solutions of the generalized
  jang equation},
        date={2013},
     journal={Comm. Partial Differential Equations},
      volume={38},
       pages={2199\ndash 2237},
}

\bib{HawkingEllis}{book}{
      author={Hawking, S.W.},
      author={Ellis, G.F.R.},
       title={The large scale structure of space-time},
   publisher={Cambridge University Press},
        date={1973},
}

\bib{Hayward}{article}{
      author={Hawyward, S.A.},
       title={Gravitational energy in spherical symmetry},
        date={1996},
     journal={Phys. Rev. D},
      volume={53},
       pages={1938\ndash 1949},
}

\bib{HuiskenIlmanen}{article}{
      author={Huisken, G.},
      author={Ilmanen, T.},
       title={The inverse mean curvature flow and the {R}iemannian {P}enrose
  inequality},
        date={2001},
     journal={J. Differential Geom.},
      volume={59},
       pages={353\ndash 437},
}

\bib{Jang}{article}{
      author={Jang, P.S.},
       title={On the positivity of energy in general relativity},
        date={1978},
     journal={J. Math. Phys.},
      volume={19},
       pages={1152\ndash 1155},
}

\bib{JaraczCylindricalPenrose}{article}{
      author={Jaracz, J.~S.},
       title={The penrose inequality and positive mass theorem with charge for
  manifolds with asymptotically cylindrical ends},
        date={2020},
     journal={Annales Henri Poincare},
      volume={21},
       pages={2581\ndash 2609},
}

\bib{JaraczWEC}{article}{
      author={Jaracz, J.S.},
       title={Spherically symmetric counter examples to the penrose inequality
  and the positive mass theorem under the assumption of the weak energy
  condition},
        date={2023},
     journal={Class. Quantum Grav.},
      volume={40},
       pages={025005},
}

\bib{KhuriSokolowskyWeinstein}{article}{
      author={Khuri, M.A.},
      author={Sokolowsky, B.},
      author={Weinstein, G.},
       title={A penrose-type inequality with angular momentum and charge for
  axisymmetric initial data},
        date={2019},
     journal={Gen. Relativity Gravitation},
      volume={51},
      number={9},
}

\bib{KhuriWeinsteinYamada1}{article}{
      author={Khuri, M.A.},
      author={Weinstein, G.},
      author={Yamada, S.},
       title={Extensions of the charged riemannian penrose inequality},
        date={2015},
     journal={Class. Quantum Grav.},
      volume={32},
       pages={035019},
}

\bib{Penrose}{article}{
      author={Penrose, R.},
       title={Naked singularities},
        date={1973},
     journal={Ann. N.Y. Acad. Sci.},
      volume={224},
       pages={125\ndash 134},
}

\bib{SchoenYau}{article}{
      author={Schoen, R.},
      author={Yau, S.T.},
       title={On the proof of the positive mass conjecture in general
  relativity},
        date={1979},
     journal={Commun. Math. Phys.},
      volume={65},
      number={1},
       pages={45\ndash 76},
}

\bib{SchoenYau2}{article}{
      author={Schoen, R.},
      author={Yau, S.T.},
       title={Proof of the positive mass theorem {II}},
        date={1981},
     journal={Commun. Math. Phys.},
      volume={79},
      number={2},
       pages={231\ndash 260},
}

\bib{Wald}{article}{
      author={Wald, R.},
       title={General {R}elativity},
        date={1984},
}

\bib{Williams}{article}{
      author={Williams, H.},
       title={A {PDE} proof of the penrose inequality for perturbations of
  schwarzschild initial data},
        date={2022},
     journal={Class. Quantum Grav.},
      volume={39},
       pages={225001},
}

\end{biblist}
\end{bibdiv}

 \footnotesize

  J.S.~Jaracz, \textsc{Department of Mathematics, Texas State University,
    San Marcos, TX 78666}\par\nopagebreak
  \textit{E-mail address} \texttt{jaracz@txstate.edu}

\end{document}